# Concept-of-Operations Disposal Analysis of Spacecraft by Gossamer Structure

Malcolm Macdonald[*], Colin M<sup>c</sup>Innes[†] and Charlotte (Lücking) Bewick[‡]
*Advanced Space Concepts Laboratory, Department of Mechanical & Aerospace Engineering,*
*University of Strathclyde, Glasgow, G1 1XQ, Scotland*

Lourens Visagie[§] and Vaios Lappas[**]
*Surrey Space Centre, Faculty of Engineering and Physical Sciences,*
*University of Surrey, Guildford, GU2 7XH, United Kingdom*

*and*

Sven Erb[††]
*ESA ESTEC, Keplerlaan 1, 2201 AZ Noordwijk, The Netherlands*

**A gossamer structure for end-of-life disposal of spacecraft to mitigate space debris is considered in comparison with other end-of-life disposal concepts to determine when it would be preferable. A needs analysis, potential use cases, and concept-of-operations are developed. A survey of disposal strategies is presented for comparison prior to a down-selection of viable competing**

[*] *Associate Director, Advanced Space Concepts Laboratory, Mechanical & Aerospace Engineering, University of Strathclyde, Glasgow, G1 1XQ, Scotland. AIAA Associate Fellow.*

[†] *Visiting Professor, Advanced Space Concepts Laboratory, Mechanical & Aerospace Engineering, University of Strathclyde, Glasgow, G1 1XQ, Scotland. AIAA member.*

[‡] *Research Assistant, Advanced Space Concepts Laboratory, Mechanical & Aerospace Engineering, University of Strathclyde, Glasgow, G1 1XQ, Scotland. Now a Space Systems and Mission Concepts Engineer at OHB System, Bremen, Germany.*

[§] *Research Assistant, Surrey Space Centre, Faculty of Engineering and Physical Sciences, University of Surrey, Guildford, GU2 7XH, United Kingdom.*

[**] *Professor, Surrey Space Centre, University of Surrey, Guildford, GU2 7XH, United Kingdom.*

[††] *GNC Systems Engineer, ESA ESTEC, Keplerlaan 1, 2201 AZ Noordwijk, The Netherlands.*

techniques; solar sailing, high and low-thrust propulsion, and electrodynamic tethers. A parametric comparison of the down-selection competing techniques is presented. Exploiting solar radiation pressure on the structure is of limited value. Atmospheric drag augmentation was found to be of most benefit for end-of-life disposal when an entirely passive means is required, allowing the gossamer device to act as a 'fail-safe'. This is applicable to only low and medium mass spacecraft, or spacecraft that are unlikely to survive atmospheric re-entry, hence minimizing risk to human life. It does not significantly alter the operating ceiling altitude but does the maximum allowable end-of-life mass. Peak mass benefit occurs in the altitude range 550 – 650 km and is largely independent of de-orbit time.

## Nomenclature

$A$ = Cross-sectional surface area, $m^2$

$A_T$ = Cross-sectional tether area, $m^2$

B = Dipole field strength, $T$

$C_D$ = Drag co-efficient

$E$ = Specific orbital energy, $J/kg$

$F_D$ = Drag force, $N$

$F_{LT}$ = Low-thrust propulsion force, $N$

$F_{srp}$ = Solar radiation force, $N$

$F_T$ = Electrodynamic tether force, $N$

$g$ = Standard gravity = 9.80665 $ms^{-2}$

$H$ = Density scale height, $km$

| $I$ | = Induced current, $A$ |
|---|---|
| $I_{sp}$ | = Specific impulse, $s$ |
| $L$ | = Lifetime duration, $s$ |
| $L_T$ | = Tether length, $m$ |
| $m_{prop}$ | = Propellant mass, $kg$ |
| $m_{sc}$ | = Spacecraft mass, $kg$ |
| $P$ | = Solar radiation pressure, $4.56 \mu\ Pa$ |
| $R_\oplus$ | = Mean volumetric radius of the Earth = 6371 km |
| $R_T$ | = Electrodynamic tether resistance, $\Omega$ |
| $r$ | = radius, $m$ |
| $v$ | = velocity, $m\ s^{-1}$ |

| $\eta$ | = geometric efficiency, a function of the solar sail steering law used |
|---|---|
| $\mu$ | = Gravitational parameter = $3.986032 \times 10^{14}\ m^3 s^{-2}$ |
| $\rho$ | = local air density, $kg\ m^{-3}$ |
| $\varphi$ | = Voltage in electrodynamic tether, $V$ |

## I. Introduction

The rapid development of space technology in the second-half of the twentieth-century led to the emergence of a shell of synthetic debris around the Earth. Within 2000 km of the Earth's surface, this shell of debris now poses a greater threat to spacecraft than the natural meteoroid environment [1, Ch. 3]. Resultantly, the Inter-Agency Space Debris Coordination Committee, IADC, have developed best-practice mitigation guidelines to limit the further generation of synthetic debris around the Earth based on three fundamental principles: limiting debris released during normal operations; minimizing the risk of on-orbit break-ups and collisions; and the focus of this paper, limiting the orbital lifetime of non-functioning objects in populated regions [2].

Spacecraft end-of-life disposal by means of drag-augmentation has been widely discussed from a technical perspective [3–7]. Drag-augmentation is typically achieved by two means, either through the deployment of a spherical envelope, for example a balloon, which has the advantage of being an omni-directional system, or by deployment of a shaped gossamer structure. Drag-augmentation concepts have also been extended to exploit enhanced solar radiation pressure on the deployed surface [8,9], and the $J_2$ perturbation [10]. The simplicity and robustness of a spherical envelope such as a balloon for de-orbiting comes with a mass penalty when compared to a shaped gossamer structure. Considering alone the material required, the surface area of a sphere goes as $4\pi r^2$, where $r$ is the characteristic length, e.g. radius, while the surface area of a flat disc or square scales as $\pi r^2$ or $r^2$, respectively.

This paper considers, in comparison with other end-of-life disposal concepts, a gossamer structure for end-of-life disposal of spacecraft to mitigate space debris to determine when it would be preferable. On this basis, a needs analysis, potential use cases, and concept-of-operations are developed. The paper is divided into four principle sections; initially a survey of re-orbiting strategies and techniques is presented to enable a down-selection of these against which to quantitatively assess the performance of the gossamer structure. The quantitative assessment is used thereafter to analyse the respective performance of each of the short-listed re-orbiting strategies and techniques against a gossamer structure in-order to determine the unique advantages of such a structure, hence determining when it would be the preferable option. Having identified this, potential use cases of a gossamer structure are developed to quantify the value of a gossamer structure in end-of-life disposal. Finally, an assessment of needs, of the gossamer structure, to realise the identified value is developed.

The gossamer structure is define *a priori* to be of order $25\text{m}^2$ projected surface area, for example a flat rectangular surface, a shaped parafoil-like surface, or a closed structure such as a balloon. The *a priori* approximate size definition is based on the volume of prior work considering structures of this scale order [6–8,11]. It is of note that the simultaneous use of multiple forces is not considered within this assessment.

## II.  Survey of Re-Orbiting Strategies and Technologies

A wide range of disposal concepts and strategies will initially be surveyed to allow selection of a range of concepts for quantitate analysis and comparison to a gossamer structure. This initial survey will be based on a range of criteria introduced in the following sections.

### A.  Orbits

To enable a detailed analysis of de-orbiting concepts, strategies and/or techniques the orbital environment is sub-divided into a number of orbit categories, as summarized in Table 1. Each de-orbiting concept, strategy and/or technique will thus be rated on its applicability within each orbital regime as **I**(*napplicable*), **L**(*ow*), **M**(*edium*) or **H**(*igh*).

| Category ID | Name | Altitude (km) | Removal Required |
|---|---|---|---|
| 1.1 | low LEO | 100 – 300 | yes |
| 1.2 | international space station region | 300 – 500 | yes |
| 1.3 | medium LEO | 500 – 1000 | yes |
| 1.4 | high LEO | 1000 – 2000 | yes |
| 2.1 | MEO | 2000 – 19000 & 24000 – 35586 | no |
| 2.2 | GNSS region | 19000 – 24000 | no |
| 3 | geosynchronous orbits | 35586 – 35986 | yes |
| 4 | supersynchronous orbits | 35956 – 45000 | no |
| 5 | HEO | all | yes |

**Table 1    Orbit categories**

## B.  Type of Method

Considering all de-orbiting concepts, strategies and/or techniques it is noted that three distinct types of method can be defined to aid the characterization, comparison and evaluation of re-orbiting technology. These types are, '*Type A*', a method applied individually to a piece of debris, either as part of the initial system development or later attached to piece of debris or a non-operational spacecraft. '*Type B*', Active Debris Removal, a method remotely applied to an individual piece of debris or a non-operational spacecraft. Or '*Type C*', a method applied universally to all objects within a certain region.

## C.  Result of Method

Considering all de-orbiting concepts, strategies and/or techniques it is noted that two distinct solutions occur; specifically these results are either a controlled or uncontrolled re-orbit or de-orbit. When large spacecraft de-orbit significant fragments can be expected to reach the Earth's surface, as witnessed when the German Roentgen Satellite (Rosat) re-entered the atmosphere on Sunday October 23 2011 over the Bay of Bengal. The uncontrolled de-orbit of Rosat ended in a 'harmless' splashdown. However, it is of note that had Rosat re-entered as little as 10 – 15 minutes later it could have impacted the Chinese mainland in the region of the cites of Chongqing and Chengdu, with a total population of over 42 million people. Therefore, it is clear that for large spacecraft a controlled de-orbit is highly desirable to ensure minimum risk to human life, typically by ensuring a splashdown in the southern Pacific Ocean 'spacecraft cemetery'. Meanwhile for small spacecraft, where complete destruction is ensured due to atmospheric heating, an uncontrolled de-orbit is acceptable. It is noted that when a spacecraft is being re-orbited then a controlled method of re-orbiting is required to avoid unforeseen collisions and subsequent legal difficulties. Each de-orbiting concept, strategy and/or technique will be rated on its result as **C**(ontrolled) or **U**(ncontrolled). In addition, each de-orbiting concept, strategy and/or technique will be rated as active or not, based on whether it would require active spacecraft operations to maintain the de-orbit concept.

## D. Comparison Metrics

Each surveyed disposal concept will be ranked **L**(*ow*), **M**(*edium*), or **H**(*igh*) against the range of metrics defined in Table 2. It should be noted from Table 2 that not all comparison metrics should aim to be low or high, for example, a high Technology Readiness Level (TRL) and low Advancement Degree of Difficulty ($AD^2$) are desired [1 Ch. 2, 12]. In addition to this, each will be categorized for insensitivity to end-of-life orbit eccentricity and inclination.

| Metric | Low | Medium | High |
|---|---|---|---|
| technology readiness level, TRL | TRL 1 – 3 | TRL 4 – 6 | TRL 7 – 9 |
| advancement degree of difficulty, $AD^2$ | $AD^2$ 1 – 3 | $AD^2$ 4 – 6 | $AD^2$ 7 – 9 |
| mass efficiency | > 15% of total mass fraction | 5 – 15% of total mass fraction | < 5 % of total mass fraction |
| volume efficiency | > 15% of total volume | 5 – 15% of total volume | < 5 % of total volume |
| sensitivity to spacecraft mass | < 50 kg | 50 – 1000 kg | 1000 kg |

**Table 2    Comparison metrics**

## E. Matrix of De-Orbiting Concepts

A wide range of de-orbiting, including re-orbiting, strategies and technologies are summarized in matrix form to allow a qualitative comparison in Table 3. Where multiple entries are given in the 'Craft Mass' column a range should be interpreted, furthermore entries in this column in bold highlight a particular efficiency. The cells are coded to ease comparison, **bold** is good, standard-text is medium, *italic* is poor, while inapplicable are grey. The columns for 'Result', 'Active' and 'Craft Mass' are not coded as favourable content within these columns is contextual and cannot be generalized.

| Name | Type | Orbit Category | | | | | | | | | Technology | | Efficiency | | Insensitivity | | Result | Active | Craft Mass |
|---|---|---|---|---|---|---|---|---|---|---|---|---|---|---|---|---|---|---|---|
| | | 1.1 | 1.2 | 1.3 | 1.4 | 2.1 | 2.2 | 3 | 4 | 5 | TRL | AD$^2$ | mass | volume | ecc. | inc. | | | |
| atmospheric drag augmentation [3–7] | A | **H** | **H** | M | *L* | I | I | I | I | I | *L* | M | M | M | M | **H** | U | No | LM |
| electrodynamic tethers [13] | A | **H** | **H** | M | *L* | I | I | I | I | I | *L* | M | M | M | M | M | U | No | LM |
| lorentz-augmented deorbiting | A | **H** | **H** | M | *L* | I | I | I | I | I | *L* | *L* | M | H | H | M | M | U | Yes | L |
| cold gas | A | **H** | M | *L* | *L* | *L* | *L* | *L* | *L* | *L* | **H** | **L** | *L* | *L* | **H** | **H** | C | Yes | LM |
| mono propellant | A | **H** | **H** | M | *L* | *L* | *L* | M | *L* | *L* | **H** | **L** | M | M | **H** | **H** | C | Yes | LM |
| bi-Propellant | A | **H** | **H** | **H** | M | *L* | *L* | **H** | *L* | *L* | **H** | **L** | M | M | **H** | **H** | C | Yes | MH |
| solid propulsion | A | **H** | **H** | M | M | *L* | *L* | M | *L* | M | **H** | **L** | M | M | **H** | **H** | C | Yes | LMH |
| hybrid propulsion | A | **H** | **H** | **H** | **H** | M | M | **H** | **H** | **H** | M | **L** | **H** | **H** | **H** | **H** | C | Yes | MH |
| electrical propulsion | A | **H** | **H** | **H** | L | L | **H** | L | **H** | M | **H** | **L** | **H** | **H** | M | **H** | U | Yes | LM |
| active solar sailing [14 Ch. 1] | A | I | I | I | L | L | L | L | M | L | M | M | M | M | M | **H** | U | Yes | LM |
| SRP on panels | A | I | I | I | L | L | L | L | M | L | **H** | **L** | **H** | **H** | M | **H** | U | Yes | LM |
| SRP-augmented deorbiting | A | I | I | L | M | M | M | M | L | M | M | M | M | M | *L* | *L* | U | No | LM |
| ground-based laser ablation [15–18] | B | **H** | **H** | M | M | *L* | I | I | I | I | *L* | *H* | **H** | **H** | M | M | U | Yes | LM |
| space-based laser ablation [15–18] | B | **H** | **H** | M | *L* | *L* | *L* | *L* | *L* | *L* | *L* | *H* | **H** | **H** | M | **H** | U | Yes | LM |
| Space-based solar ablation [15–18] | B | **H** | **H** | M | *L* | *L* | *L* | *L* | *L* | *L* | *L* | *H* | **H** | **H** | M | **H** | U | Yes | LM |
| multi-layered sphere | B | **H** | **H** | M | *L* | *L* | *L* | *L* | *L* | *L* | M | *H* | **H** | **H** | M | **H** | U | No | L |
| foam-based ADR [19] | B | **H** | **H** | M | I | I | I | I | I | I | *L* | *L* | **H** | **H** | M | **H** | U | Yes | L |
| ion-beam shepherd [20] | B | **H** | **H** | **H** | M | *L* | *L* | **H** | *L* | M | M | M | **H** | **H** | **H** | **H** | C | Yes | LMH |
| space tug [21] | B | **H** | **H** | **H** | M | *L* | *L* | **H** | *L* | M | M | M | **H** | **H** | **H** | **H** | C | Yes | MH |
| drag | C | **H** | M | *L* | I | I | I | I | I | I | **H** | **L** | n/a | n/a | M | **H** | U | No | LM |
| catcher's mitt [22] | C | **H** | I | M | *L* | *L* | *L* | I | *L* | *L* | *L* | *H* | n/a | n/a | M | **H** | U | Yes | L |
| tungsten dust [23,24] | C | *L* | M | **H** | *L* | I | I | *L* | *L* | *L* | *L* | *H* | n/a | n/a | M | **H** | U | No | LM |

**Table 3    De-orbiting, including re-orbiting, strategies and technologies summarized in matrix form**

## F.  Initial Down-Selection of Concepts

At this initial stage, a down-selection can be performed on evidently unviable strategies (in the 10-15 year timeframe), or strategies that only apply to regions where a gossamer structure would not be applicable. As such, all strategies and/or technologies which have a **L**(*ow*) TRL or **H**(*igh*) AD$^2$ are removed from further consideration at this point.

A gossamer structure can have two modes of operation. These are, where atmospheric drag is used to de-orbit a satellite, or where solar radiation pressure is used to either re-orbit a spacecraft at the end-of-life, for example in Geostationary Orbit, GEO, or to lower a satellite's orbit such that atmospheric drag can be used to complete the de-orbit process. It is apparent however given the high end-of-life mass (>3000 kg) of the vast majority of spacecraft in GEO that the use of a gossamer structure, of order 25 m$^2$, for solar sailing to re-orbit a spacecraft from GEO to the graveyard/supersynchronous region is significantly sub-optimal when compared against a conventional end-of-life bi-propellant manoeuvre. Furthermore, as the magnitude of propulsive force from a gossamer structure would be small it is likely that the increase in orbit altitude per orbit would be insufficient to prevent the spacecraft entering the neighbouring spacecraft slots. Considering this it is apparent from Table 3 that the remaining, and hence down-selected concepts for further quantitate analysis are limited to the LEO region, that is up to 2000 km, where the gossamer structure would be used to augment the atmospheric drag effects or as a solar sail to enter the region of atmospheric drag. The down-selected concepts are: atmospheric drag augmentation; electrodynamic tethers; mono-propellant; bi-propellant; low-thrust, high specific impulse propulsion; and, solar sailing to gain atmospheric drag. Note that Cold Gas propulsion was not down-selected due to its inefficiency, whilst solid propulsion was not down-selected due to the risk of creating further debris from propellant slag.

### III.  Comparison of a Gossamer Structure against Down-Selected Concepts

To provide a quantitative assessment of the performance of competing de-orbit methods, a series of analytic approximations are developed that provide approximate requirements for the de-orbit device to ensure de-orbit within a fixed duration. The de-orbit timescale $L$ is fixed at 25 years in all cases to comply with the IADC space debris mitigation guidelines. The analysis assumes a quasi-circular low thrust spiral affected by air drag, solar radiation pressure or low-thrust electric propulsion. For air drag de-orbit it is assumed that the drag device is stabilized and is always normal to the velocity vector. For the solar sail, two steering laws are required, either for near-equatorial or near-polar orbits.

### A.  De-Orbit Scaling Laws

#### 1.  Drag Augmentation

The decay time-scale for a drag augmented device can be estimated by considering the work done by the drag force $F_D$ on a spacecraft of mass $m_{sc}$ and total drag cross-sectional area $A$ on a circular orbit of radius $r$. Assuming a quasi-circular orbit the spacecraft orbit speed is therefore $v = \sqrt{\mu/r}$, where $\mu$ is the gravitational parameter. The rate of change of two-body specific energy $E = -\mu/2r$ is then given by

$$\dot{E} \cong -\frac{1}{m_{sc}} F_D v \tag{1}$$

where the drag force $F_D$ is defined as

$$F_D = \frac{1}{2} C_D \, A \, \rho \, v^2 \tag{2}$$

with $C_D$ the drag coefficient (assumed to be 2.1) and $\rho$ the local air density. In order to proceed, an analytic model of the atmospheric density is required. Using a power law fit to the 1976 standard atmosphere [25] from 150-1000 km it is found that

$$\rho (kg\ m^{-3}) = \Lambda h(km)^{-\gamma} \tag{3}$$

where h is height, $\Lambda = 10^7$ and $\gamma = 7.201$; giving a coefficient of determination, $R^2$, of 0.998. Note that the 1976 standard atmosphere is a static model and does not account for variability in the atmosphere, including the impact of the solar cycle, rather it provides a mean atmosphere. As the de-orbit timescale $L$ is fixed at 25 years it is reasonable to average the effects due to the approximately 11 year solar cycle.

Integrating along the quasi-circular orbit decay spiral from some initial orbit radius $r_0$ to some final orbit radius $r_1$ for some duration $L$ it can be shown that the required drag area is given by

$$A = \frac{m_{sc}}{C_D\ L\ \Lambda \sqrt{\mu\ R_\oplus}} \frac{((r_0 - R_\oplus)^{1+\gamma} - (r_1 - R_\oplus)^{1+\gamma})}{1+\gamma}\ 1000^{-\gamma} \tag{4}$$

where $R_\oplus$ is the mean volumetric radius of the Earth (6371 km) and all distances are given in meters. For a given spacecraft mass and initial orbit, the total drag area can then be estimated for a fixed de-orbit duration, again assumed to be 25 years.

## 2. Solar Sailing

Solar sailing is assumed to be inapplicable for orbits below 750 km altitude, due to the dominance of atmospheric drag in this region [14,26]. However, the orbit transfer time for a solar sail can be estimated using a similar analysis to that used in the previous section, with an appropriate sail steering law. For equatorial orbits, a simple switching 'on/off' control law can be used that requires a sail slew of 90 degrees twice per orbit. For polar orbits, the sail attitude can be fixed relative to the Sun, but the sail must yaw 360 degrees per orbit to align the sail thrust vector with the velocity vector [14 Ch. 4].

The rate of change of two-body specific energy $E = -\mu/2r$ can be obtained from the work done by the solar sail generated thrust $F_{srp}$ such that

$$\dot{E} \cong -\frac{1}{m_{sc}} F_{srp} v \ . \tag{5}$$

Therefore, the rate of change of orbit radius can be found from

$$\frac{\mu}{2r^2} \frac{dr}{dt} \approx \frac{F_{srp}}{m_{sc}} \sqrt{\frac{\mu}{r}} \tag{6}$$

which can be integrated to provide the orbit radius as a function of time. Assuming an ideal reflector, the thrust $F_{srp}$ is related to the sail area $A$ and the solar radiation pressure $P$ by

$$A = \frac{F_{srp}}{2P} \ . \tag{7}$$

For a continuous, quasi-circular low thrust spiral from some initial orbit radius $r_0$ to some final orbit radius $r_1$ in duration $L$ it can be shown that the required thrust-to-mass ratio is given by

$$A = \eta \frac{m_{sc}}{2 \, P \, L} \left| \sqrt{\frac{\mu}{r_1}} - \sqrt{\frac{\mu}{r_0}} \right| \tag{8}$$

where $\eta$ is the geometric efficiency, a function of the sail steering law used. Effectively, the geometric efficiency gives a relationship between the averaged orbital transverse acceleration over one orbit revolution for each sail steering law, and is discussed in detail in [14 Ch. 4].

For near-equatorial orbits, a simple switching 'on-off' control law is used that requires a sail slew of 90° twice per orbit. It can be shown that the steering law geometric efficiency is $\eta = \pi$, which accounts for the loss of thrust due to the switching law [14 Ch. 4]. For near-polar orbits, the sail attitude is fixed relative to the Sun and yaws 360 degrees per orbit to align the sail thrust and velocity vector's. Here the steering law geometric efficiency is $\eta = 2.83$, which accounts for the loss of thrust due to the pitch of the sail relative to the Sun-line [14 Ch. 4]. It can be seen that the polar orbit steering law is marginally more efficient than the equatorial steering law. Note that the effect of shadow is neglected in this analysis.

### 3. Low-Thrust Propulsion

Using a similar analysis to previous sections, the decay time for continuous low thrust electric propulsion can be determined. The thrust $F_{LT}$ required to de-orbit from some initial orbit radius $r_0$ to some final orbit radius $r_1$ in some fixed duration $L$ is found to be

$$F = \frac{m_{sc}}{L} \left| \sqrt{\frac{\mu}{r_0}} - \sqrt{\frac{\mu}{r_1}} \right|. \tag{9}$$

The required propellant mass can then be determined from the effective $\Delta v$ such that

$$m_{prop} = m_{sc}\left(1 - e^{-\Delta v / g\, I_{sp}}\right) \tag{10}$$

where $I_{sp}$ is the specific impulse of the propulsion system, assumed to be a conservative 3000 seconds throughout this paper, and

$$\Delta v = \left| \sqrt{\frac{\mu}{r_0}} - \sqrt{\frac{\mu}{r_1}} \right|. \tag{11}$$

### 4. Electrodynamic Tether

To account for the variation in tether attitude, it is assumed that the tether is aligned off the local vertical at a mean angle of $\tan^{-1}\left(1/\sqrt{2}\right)$ and is kept in tension. A conducting tether of length $L_T$ moving at speed $v$ through a magnetic field of strength $B$ has an induced voltage $\varphi = L_T v B$, where $B$ is the dipole field strength given by

$$\text{B} = B_o \left(\frac{R_\oplus}{r}\right)^3 \tag{12}$$

for field strength $B_o = 3.5\text{x}10^{-5}\ T$. The force induced in the tether is then $F_T = BIL_T$, where $I$ is the current induced by the voltage $\varphi$. The current and voltage are related by Ohms law such that $\varphi = IR_T$ for tether resistance $R_T$, which in principle includes the resistance of the plasma contactors and the plasma sheath. Assuming that the tether follows a quasi-circular inward spiral, the speed $v$ is then approximated by the local circular orbit speed $v = \sqrt{\mu/r}$. The force acting on the tether can therefore be estimated as

$$\text{F}_T = B_o^2 \left(\frac{R_\oplus}{r}\right)^6 \frac{L_T^2}{R_T} \sqrt{\frac{\mu}{r}}. \tag{13}$$

The tether resistance $R_T$ can be scaled with tether length through resistivity $\rho$, such that $R_T = \rho\, L_T/A_T$, where the resistivity of aluminum is assumed to be $\rho \approx 2.82\text{x}10^{-8}$ Ohm-m and $A_T$ is the cross-sectional area of the tether. The tether is assumed to have a diameter of 2 mm.

Finally, in order to model the interaction of a tether in an inclined orbit with the magnetic field a geometric efficiency factor $\varepsilon$ is defined following [27] as

$$\varepsilon = \frac{1}{16}\left\{\cos\left[\tan^{-1}\left(1/\sqrt{2}\right)\right]\right\}^2 \{6 + 2\cos 2i + 3\cos[2(i - \delta)] + 2\cos 2\delta \tag{14}$$
$$+\, 3\cos[2(i + \delta)]\}$$

where $\delta$ is the tilt of the Earth's dipole field relative to the equator, assumed to equal 11.5 degrees. Using Eq. (6), and integrating, the tether length $L_T$ required to de-orbit in duration $L$ is given by

$$L_T = \frac{(r_1^6 - r_o^6)}{12\ \varepsilon\ L} \frac{m_{sc}}{R_E^6 B_o^2} \frac{\rho}{A_T} \tag{15}$$

from some initial orbit radius $r_0$ to some final orbit radius $r_1$.

**B.  Analysis**

Considering the Iridium (781 km altitude at 86.4° inclination), GlobalStar (1410 km altitude at 52° inclination) and Orbcomm (825 km altitude at 52° inclination) communication constellations as test cases, each of the down-selected concepts can be analytically considered in-turn to quantify the relative value of each.

**1.  De-Orbit by Atmospheric Drag Augmentation**

The analytic scaling laws can be used to determine the required drag area required, as shown in Table 4. Note that the power law fit in Eq. (3) is assumed to remain valid up to the altitude of the GlobalStar spacecraft. For the purposes of analysis the de-orbit time $L$ is fixed as 25-years and de-orbit is assumed at an altitude of 100 km.

|  | Iridium | GlobalStar | Orbcomm |
|---|---|---|---|
| initial altitude (km) | 781 | 1410 | 825 |
| bus mass (kg) EOL | 526 | 546 | 100 |
| bus area (m$^2$) low {high} | 4 {12} | 4 | 1 |
| required drag area (m$^2$) | 40.6 | 5351 | 12 |
| equivalent square side-length (m) | 6.4 | 73.2 | 3.5 |

Table 4    Required total drag area to de-orbit in 25 years (drag augmentation effects only)

Note that, assuming an Iridium spacecraft surface area of 4 and 12 square meters respectively, corresponding to spacecraft body only and spacecraft body plus maximum solar array area, it is found that an Iridium spacecraft will naturally decay in approximately 90 – 250 years. However, reducing the initial circular orbit altitude to 575 km, this natural decay timescale reduces to between 7 – 21 years. Similarly, a GlobalStar spacecraft can be assumed to, in-effect, never decay due to atmospheric drag alone, however reducing its altitude to 575 – 585 km altitude reduces the decay time to atmospheric drag to approximately 21 – 24.5 years. Finally, an Orbcomm spacecraft can be assumed to decay due to atmospheric drag alone in approximately 300 years, however reducing its altitude to 600 – 605 km reduces the decay time to atmospheric drag to approximately 22 – 24 years. The effect of such limited use of propulsion will be discussed later.

## 2. De-Orbit by Electrodynamics Tether

The analytic scaling laws for the electrodynamic tether can be used to determine the required tether length. For a de-orbit time $L$ of 25-years, with de-orbit at an altitude of 100 km, it is found that the required tether lengths are short; a direct result of the 25-year de-orbit timeframe. Most prior electrodynamic tether de-orbit analysis has sought to complete the de-orbit in a much shorter period, for example in [28] de-orbit times of 7.5 months are sought for tethers of up to 20 km. Operationally there is no 'strict' need to de-orbit in less than 25-years as once deployed the electrodynamic tether should remain operationally passive, assuming sufficient on-board autonomy. However, the electrodynamic tether is not an inert system, as it requires an active electrical control system to remain operational. Therefore, it is likely that for reliability reasons a shorter de-orbit would be preferred. Reducing the de-orbit time to 5 % of the operational life of the spacecraft, that is ¾ of a year and 100-days assuming an operational lifetime between 5 and 15 years, the required tether lengths are found to vary from: 0.8 – 2.1 km (Iridium); 0.1 – 0.3 km (GlobalStar); and, 10 – 30 m (Orbcomm). Note that the lower inclination of the GlobalStar spacecraft result in a reduced tether length in comparison to the lower altitude, but near-polar orbiting Iridium spacecraft.

Given the requirements for non-passive, autonomous systems, it is reasonable to assume that short de-orbit times will be required to reduce the risk of system failure and to reduce the risk to other spacecraft. This coupled with the difficulties in deploying a long tether, especially from an inactive and potentially tumbling spacecraft make electrodynamic tethers a challenging concept.

## 3. De-Orbit by Low-Thrust, High-Specific Impulse Propulsion

The requirement to extend spacecraft operations for a further 25-years after the operational end-of-life is unattractive due to both the operations costs and the lifetime of most electric thrusters. It is noted that, for example, a GlobalStar spacecraft can be de-orbited, assuming continuous control, in approximately 100-days with a thrust magnitude of less than 45 mN. This notional period of extended operations can be further reduced by noting that the thrust must only be applied until the spacecraft reaches an orbit from which it will naturally decay within 25-years of the end-of-life due to atmospheric drag, as previously discussed. This hybrid method was how Iridium 9, which failed and was replaced by Iridium 84, was de-orbited; the on-board resistojets (low-thrust / low specific impulse: 300 mN / 350 seconds) were used to lower the spacecraft perigee in mid-late September 2000 and by early November the spacecraft was observed to be tumbling out of control. The orbit of Iridium 9 finally decayed in March 2003. Using the previous analysis that determined the maximum altitude for decay within <25-years due to atmospheric drag alone for each use-case spacecraft, the required propellant mass saving of such a hybrid scheme is shown in Table 5 along with the fuel mass required to directly de-orbit.

|  | Iridium | GlobalStar | Orbcomm |
|---|---|---|---|
| initial altitude (km) | 781 | 1410 | 825 |
| target altitude (km) | 575 | 575 | 600 |
| bus mass (kg) EOL | 526 | 546 | 100 |
| required fuel; direct de-orbit (kg) | 6.8 | 12.7 | 1.4 |
| required fuel; drag assisted de-orbit (kg) | 2.0 | 7.8 | 0.4 |
| fuel saving (%) | 71 | 39 | 71 |

**Table 5**     **Required propellant to de-orbit directly or to move to an orbit from which natural decay within 25-years of the end-of-life due to atmospheric drag will occur (low-thrust only)**

Note that the equivalent resistojet propelled hybrid de-orbit technique for an Iridium spacecraft reduces the propellant mass required from approximately 62 kg to approximately 17 kg. If any low-thrust propulsion system is already on-board a spacecraft in LEO it is likely an attractive option for de-orbiting an operational spacecraft at end-of-life.

### 4. De-Orbit by Mono and Bi-propellant

The required mono- or bi-propellant mass can be determined, using Eq. (10), based on the required velocity change to reduce the orbit perigee to 100 km. Of the initially down-selected concepts, this is the only one that will provide a controlled atmospheric re-entry. However, if this is not required then as in the low-thrust scenario the required propellant mass can be reduced by simply moving the spacecraft to an orbit from which it will naturally decay within 25-years of the end-of-life due to atmospheric drag. As any gossamer structure will similarly result in an uncontrolled re-entry this is the only pertinent comparison. The required propellant mass saving of such a hybrid scheme is shown in Table 6.

|  | Iridium | GlobalStar | Orbcomm |
|---|---|---|---|
| initial altitude (km) | 781 | 1410 | 825 |
| target altitude (km) | 575 | 575 | 600 |
| bus mass (kg) EOL | 526 | 546 | 100 |
| required mono-prop fuel (kg) | 26 – 41 | 113 – 179 | 5.5 – 8.4 |
| required bi-prop fuel (kg) | 14 – 20 | 58 – 83 | 2.9 – 4.1 |
| fuel saving (%) | 42 – 44 | - (26 – 29) | 41 – 44 |

**Table 6     Required propellant to move each spacecraft to an orbit from which it will naturally decay within 25-years of the end-of-life due to atmospheric drag (mono & bi-prop only)**

It is noted from Table 6 that the GlobalStar spacecraft would actually require an increased quantity of fuel in this hybrid de-orbit scenario due to the use of a circular target orbit for the manoeuvre. In-order to reduce the fuel requirement for the GlobalStar spacecraft an eccentric intermediate orbit would be required, with a perigee altitude of, perhaps, 300 – 400 km, reducing the required fuel mass by 5 – 25 % from the direct de-orbit scenario. It should also be noted that an eccentric intermediate orbit might provide further propellant mass savings for the Iridium and Orbcomm platforms however, the purpose of this analysis is to determine mass values that are sufficient for comparison rather than to fully optimise such a manoeuvre. Thus, whilst further fuel mass saving might be possible even without them if any high-thrust propulsion system is already on-board a spacecraft in LEO it is likely an attractive option for de-orbiting an operational spacecraft at end-of-life. This could be either directly, or by moving the spacecraft to an orbit from which it will naturally decay within 25-years of the end-of-life due to atmospheric drag.

## 5. Active Solar Sailing to Gain Atmospheric Drag

Assuming zero atmospheric drag when altitude is greater than 750 km, and zero solar radiation pressure when altitude is less than 750 km and noting that $L = L_{sail} + L_{drag}$, Eqs. (4) and Eq. (8) may be equated to eliminate $L_{drag}$ as it can be assumed that the surface area of the solar sail equals the surface area of the drag device, giving

$$L_{sail} = \frac{\left(-\Psi L \sqrt{\mu R_\oplus}\sqrt{\frac{\mu}{r_0}}\eta\Lambda + \Psi L \sqrt{\mu R_\oplus}\sqrt{\frac{\mu}{r_i}}\eta\Lambda - \Psi L \sqrt{\mu R_\oplus}\sqrt{\frac{\mu}{r_0}}\gamma\eta\Lambda + \Psi L \sqrt{\mu R_\oplus}\sqrt{\frac{\mu}{r_i}}\gamma\eta\Lambda\right)}{20P[(r_1 - R_\oplus)^\gamma(R_\oplus - r_1) - (r_i - R_\oplus)^\gamma(R_\oplus + r_i)] - \Psi\mu\sqrt{R_\oplus}\eta\Lambda\left[\sqrt{\frac{1}{r_0}} + \sqrt{\frac{1}{r_i}} - \gamma\left(\sqrt{\frac{1}{r_0}} + \sqrt{\frac{1}{r_i}}\right)\right]} \tag{16}$$

where, $\Psi = 21 \times 10^{3\gamma}$, $r_i$ is the radius at which the switch from solar sailing to drag occurs (equivalent to 750 km altitude) and where all radii are given in meters.

For the purposes of analysis the de-orbit time $L$ is fixed as 25-years and de-orbit is assumed at an altitude of 100 km. The required solar sail / drag surface area can thus be determined for de-orbit within the required total period: 33 m$^2$ (Iridium); 100 m$^2$ (GlobalStar); 7 m$^2$ (Orbcomm). The length of time for which spacecraft operations must be extended, during the solar sailing phase, assuming passive stabilization during drag augmentation and a polar orbit can also be determined: 3 years (Iridium); 17 years (GlobalStar); 5 years (Orbcomm). It is found that using solar sailing to gain atmospheric drag can be immediately discarded due to the prolonged period of extended spacecraft operations required to deliver a meaningful effect. Whilst the period of extended spacecraft operations could be reduced, it would have a direct and significant impact on the required size of the gossamer structure, or the mass of the spacecraft that could be de-orbited.

## 6. Re-Orbit by Low-Thrust, High-Specific Impulse Propulsion

It is of interest to consider re-orbiting space objects rather than de-orbiting them as this may require a reduced amount of energy input. Furthermore, the convention on international liability for damage caused by space objects, the Liability Convention [29], states in Article II that a launching state shall be absolutely liable to pay compensation for damage caused by its space objects on the surface of the Earth, or to aircraft in flight. While Article III states that, the launching state shall be liable only if damage caused in space "*is due to its fault or the fault of persons for whom it is responsible.*" Therefore, the reduced level of liability adds further value to considering a re-orbit manoeuvre to a graveyard orbit.

The analytic scaling laws can be used to determine the required fuel mass to re-orbit a spacecraft above the LEO region, an altitude of >2000 km: 10 kg (Iridium); 5 kg (GlobalStar); 2 kg (Orbcomm). It is also noted that the required fuel mass to re-orbit a GlobalStar spacecraft (from 1410 km altitude) is lower than the equivalent de-orbit concepts. Hence, if an electric propulsion system is already on-board a spacecraft in LEO it is likely an attractive option for re-orbiting spacecraft at end-of-life due to both the reduced, or similar, fuel mass and the reduced level of liability.

### 7. Re-Orbit by Mono and Bi-propellant

Using Eq. (10) the required mono- or bi-propellant mass can be determined based on the required velocity change to increase the circular orbit altitude to 2000 km. It is found that both the Iridium and Orbcomm spacecraft require additional fuel to re-orbit than to de-orbit, however as previously discussed this does not rule-out such an end of life strategy due to the increased liabilities when de-orbiting. It is found that the propellant requirement for a GlobalStar to reach the graveyard region is similar to that required to move to an orbit from which it will naturally decay within 25-years of the end-of-life due to atmospheric drag. Unlike the de-orbit scenario it is not possible to reduce the propellant requirement as the spacecraft must perform two manoeuvre's to attain the final orbit.

### 8. Active Solar Sailing

The analytic scaling laws can be used to determine the required sail surface area for a 25-year re-orbit of a spacecraft to above 2000 km for a polar orbit: 117 $m^2$ (Iridium); 55 $m^2$ (GlobalStar); 21 $m^2$ (Orbcomm). In the case of a solar sail, the requirement to extend operations by 25-years after end-of-life can only be relaxed by increasing the sail size. For example, a GlobalStar re-orbit in 12 years requires a sail of approximately 145 $m^2$, while a re-orbit in 1.5 years (10 % of the spacecraft operational life) requires a sail of approximately 900 $m^2$. Furthermore, note that an Orbcomm re-orbit in 1.5 years requires a similarly large square sail of approximately 360 $m^2$.

### 9. Conclusions of Analysis

Of the initially down-selected concepts, the only one that does not appear to offer any added value is active solar sailing; principally due to the time required and the associated cost of extending spacecraft operations during this period. However, if the available solar radiation pressure could be exploited passively then it would overcome this cost issue.

If a spacecraft already has a propulsion system on-board, high or low-thrust, it is likely that this will offer an efficient de- or re-orbit strategy in comparison to a gossamer structure. This is especially so when coupled with the hybridization concept of simply manoeuvring the spacecraft to an orbit from which it will naturally decay within 25-years of the end-of-life due to atmospheric drag.

Within the medium and high LEO environment for most telecommunications constellations (or similar mass spacecraft), the use of atmospheric drag augmentation or an electrodynamic tether is of limited value when compared to the use of other technologies already on-board the spacecraft, i.e. the on-board propulsion system. However, the dual-use of such on-board technology to de-orbit the spacecraft requires that the spacecraft remains operational at the end-of-life, as indeed does an electrodynamic tether. If a spacecraft suffers a catastrophic failure, the use of on-board technology to de-orbit the spacecraft is not possible. However, an atmospheric drag augmentation system could be added to spacecraft as a 'fail-safe' de-orbiting technology, which if the spacecraft suffers a catastrophic failure would activate. The principal advantage of an atmospheric drag augmentation system, such as a gossamer structure, is therefore the entirely passive operational mode and an ability to passively maximize the surface area to the mean-free flow of the atmosphere in all of the flow regimes in which it will pass through. The use of an atmospheric drag augmentation system is applicable to only low and medium mass spacecraft, or spacecraft that are unlikely to survive atmospheric re-entry, hence minimizing risk to human life.

## IV. Potential Use Cases

To further quantify the applicability of drag augmentation methods, a more detailed analytical model of orbit decay under atmospheric drag is applied. Note that this model was not used previously such as maximise the similarity of the comparison between concepts in the previous section. This enhanced model allows for the effects of atmospheric rotation, and hence initial orbit inclination, to be included in the analysis but still assumes an initially circular orbit. It should be noted that more detailed solutions are available for eccentric orbits [30]. Using this enhanced model the approximate decay time of a circular orbit under atmospheric drag [31 p.150-151] is

$$L = \frac{1}{4\pi} \left( \frac{2\beta r_c + 1}{\rho_c \beta^2 r_c^3} \right) \left( F^{-3/2} \right) \left( \frac{m}{AC_D} \right) \left( 1 - e^{\beta\Delta} \right) \tag{17}$$

where subscript $c$ denotes the initial conditions of the circular orbit, $\Delta$ is the negative change in orbit radius, $\rho_c$ is the initial atmospheric density, assuming an exponential atmospheric model from Eq. (19), $\beta$ is defined in Eq. (20), and $F$ is defined as

$$F = \left( 1 - \frac{r_p}{v_p} \omega_{atm} \cos i \right)^2 \tag{18}$$

where $\omega_{atm}$ is the mean angular rate of rotation of the atmosphere, which is typically taken to be the same as the Earth's mean rotation rate but can vary between 0.8 and 1.3 revolutions per day. The exponential atmospheric model assumes that the atmospheric density varies as

$$\rho = \rho_0 \exp\left( -\frac{r - r_0}{H} \right) \tag{19}$$

where $H$ is the density scale height and is, for a constant $h$,

$$\beta = \frac{1}{H} = \left( \frac{1}{h} - \frac{2}{r_0} \right). \tag{20}$$

Note that the assumption of a constant scale height requires that the density scale height is also constant. While these equations are not exact, they remain valid to an altitude of several hundred kilometres. It is found that the effect of varying inclination is negligible, less than the uncertainty due to $C_D$, hence an inclination of 90-degrees is assumed in future analysis.

Using Eq. (17), the required ballistic co-efficient, $m_{sc}/(A\,C_D)$, to de-orbit in 25-years from a range of altitudes can be determined. Thereafter, assuming a surface area of 25 m$^2$ the maximum mass that can be de-orbited in 25-years across the range of altitudes can be determined, assuming a passively stable attitude is maintained to maximize the surface area exposed to the atmospheric free-stream flow.

Having determined the maximum mass that can be de-orbited with a given size of gossamer structure it is noted that the 'typical' surface area [32 p.337] of a spacecraft is

$$A = \left( \frac{\sqrt[3]{m_{sc}}}{4} \right)^2 . \tag{21}$$

Thereafter the typical surface area of a spacecraft of the maximum mass that will de-orbited in 25-yrs can be determined. That is, the surface area without the gossamer structure. Note that for low-altitude orbits, the maximum mass that will de-orbit in 25-yrs will be very large, hence it can be expected that the 'typical' surface area of such a spacecraft may be large, and may indeed be larger than the assumed gossamer surface area. Having determined the typical surface area, the allowable mass of such a spacecraft can also be determined from the known ballistic co-efficient limit for a 25-year deorbit. Hence, the mass benefit of the drag augmentation device determined. However, in determining the benefit of drag augmentation device it must be noted that not only does this change the spacecraft surface area but it also changes the drag co-efficient; the drag co-efficient is however assumed constant and low, $C_D = 2$, to allow a conservative estimate to be made.

The mass or altitude benefit of a gossamer structure projecting an area of 25 m$^2$ into the free-stream direction, for a range of de-orbit times from an orbit inclination of 90-degrees, is shown in Fig. 1. It is seen that, for example, the structure will allow a 1000 kg spacecraft to increase its operating ceiling altitude by approximately 50 km and still de-orbit within 25-years of the end-of-life. Alternatively, a spacecraft operating at 650 km altitude can increase its end-of-life mass by 736 kg (or approximately 200 %) and still de-orbit within 25-years of the end-of-life. From Fig. 1 it is seen that drag augmentation does not significantly alter the operating ceiling altitude of a spacecraft, however it does significantly alter the maximum allowable end-of-life mass.

Interpreting the data in Fig. 1, the mass benefit of drag augmentation against altitude can be further elucidated, as shown in Fig. 2 where it is seen that the peak mass benefit occurs in the altitude range 550 – 650 km and is largely independent of de-orbit time. Although Fig. 2 shows that the percentage benefit of drag augmentation increases as altitude is increased, it must be noted that this increasing altitude is a percentage of an ever-decreasing number; hence, the 'real' value of drag augmentation for the 'typical' spacecraft is limited for altitudes beyond approximately 800 km. From Fig. 2 it is apparent that drag augmentation is of little value for 'typical' spacecraft at altitudes below 550 km, as such 'typical' spacecraft can be expected to de-orbit naturally due to atmospheric drag below this altitude within the required timeframe without the need for drag augmentation.

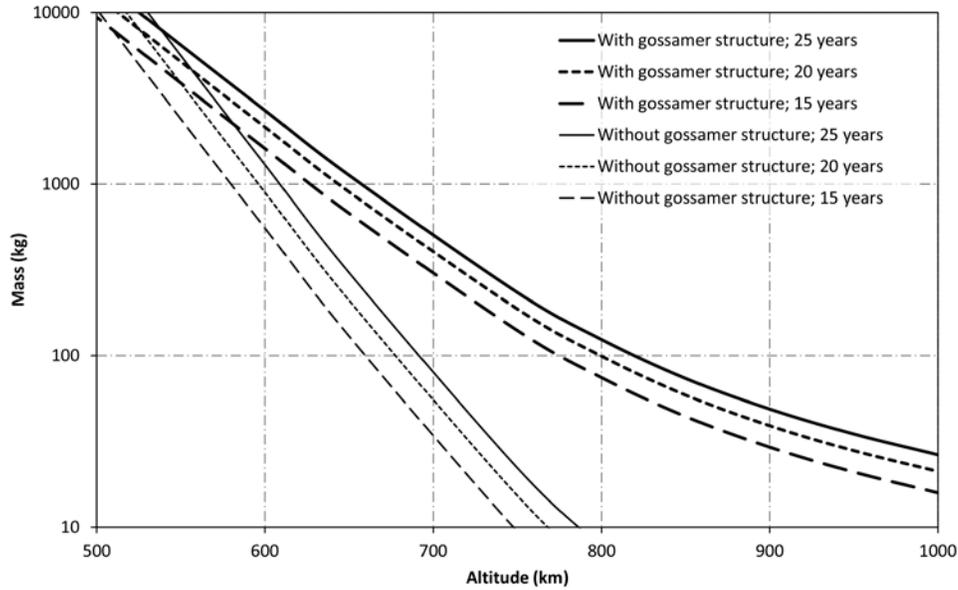

**Fig. 1   Maximum de-orbit mass for a range of upper time limits from a range of orbits, with and without a gossamer structure; 90 degree inclined orbit**

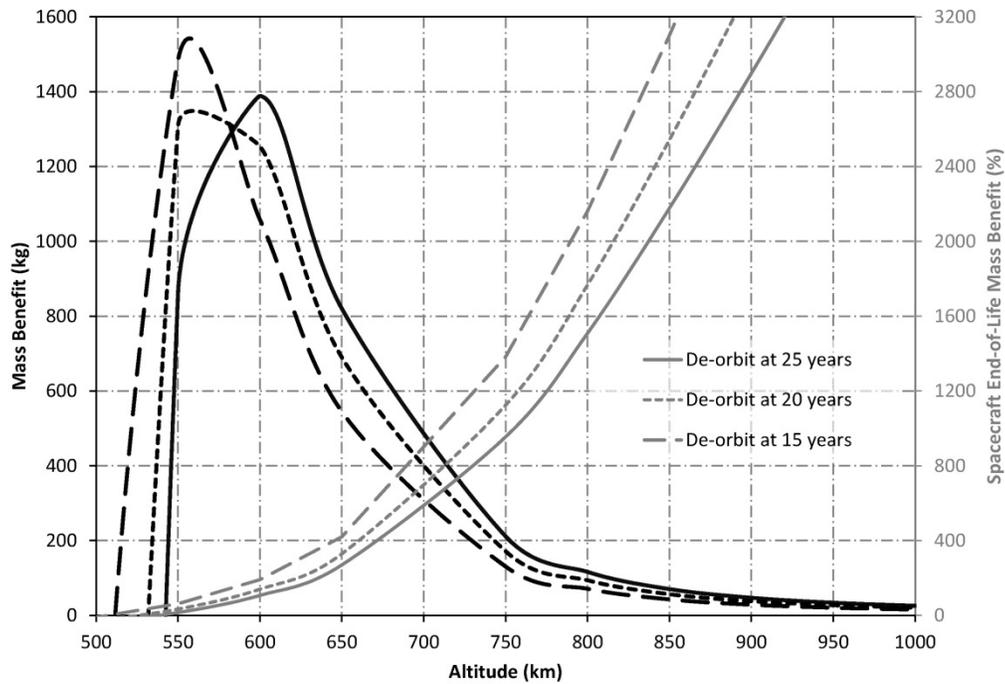

**Fig. 2   Mass benefit (and percentage benefit) in the use of a gossamer structure for de-orbiting within a range of upper time limits from a range of orbits; 90 degree inclined orbit**

Changing the surface area of the drag augmentation device allows increased masses to be de-orbited and/or allows drag augmentation to be applied at increased altitudes. Once again, the benefit of drag augmentation is considered rather than the actual mass or altitude variation. It is found that the lower altitude at which drag augmentation is of value to the 'typical' spacecraft decreases as the drag surface area is increased and that, in-general, larger surfaces can operate in a larger altitude range. Quantitatively, it is also noted that additional benefit is provided by drag augmentation when the required de-orbit time is reduced.

## A. Spacecraft Population

Considering active spacecraft in LEO in altitude bins of ±50 km of the mean altitude, it is found that two distinct peaks in population occur, in the 800 km altitude bin and in the 1400 & 1500 km altitude bin. Secondary peaks in population exist either side of the 800 km altitude bin at 600 & 700 km and 900 & 1000 km altitude.

The gossamer concept is of little value to the higher altitude peak in population. However, the concept, in its drag augmentation mode of operation, is of some value in the 800 km altitude bin and in the secondary peak population density below this bin. However, it must be recalled that the 800 km altitude population peak is towards the top-end of the useful regime for drag augmentation and depending on spacecraft size, may require a surface area significantly larger than 25 m$^2$. Beyond 800 km altitude, it is assumed drag augmentation is, at best, of moderate to poor value, however this is dependent on spacecraft mass.

## V. Assessment of Needs

The 'Outer Space Treaty' states that ownership of objects launched into space is not affected by their presence in space [33 Article VIII], while as previously discussed the 'Liability Convention' defines liability for damage caused [29]. Hence, ownership and liability for damage that results from space debris remains with the spacecraft launching state, or its designated licensed operator. However, the use of an on-board propulsion system to de-orbit a spacecraft requires that the spacecraft is operational, although perhaps only in a degraded state. If a spacecraft suffers a catastrophic failure, the use of on-board systems to de-orbit the spacecraft is not possible; however, the spacecraft launching state, or its designated licensed operator, remains liable for the risk and as such a 'fail-safe' de-orbit system may be desired to mitigate the risk of fault liability. Of all the de-orbit concepts considered following the initial down-selection only drag augmentation offers the potential to be deployed from an inactive, tumbling spacecraft.

A Summary of the needs of a gossamer structure for end-of-life disposal of spacecraft is given in Table 7, and a detailed discussion of these can be found in [34].

| Number | Need |
|--------|------|
| 1 | fully passive and require no ground operations support |
| 2 | passively aerodynamically stable in an attitude that maximizes the possible surface area to the mean-free flow of the atmosphere in all flow regimes |
| 3 | able to deploy from a tumbling spacecraft |
| 4 | able to minimize the initial angle of attack during deployment to the mean-free flow of the atmosphere |
| 5 | able to deploy from an inactive, non-responsive spacecraft |
| 6 | able to deploy autonomously at a predetermined local solar time |
| 7 | shall not increase the likelihood the spacecraft impacting other spacecraft during the de-orbit phase |
| 8 | an integral part of the spacecraft systems design from Phase 0 onwards |
| 9 | able to deploy reliably and verifiably at the end of the spacecraft's operational life |
| 10 | able to avoid an unsolicited deployment |
| 11 | able to deploy without hindrance from spacecraft peripherals such as antenna or solar arrays |
| 12 | not adversely alter the heating load profile during re-entry, such as to reduce the amount of material destroyed during re-entry |
| 13 | de-orbit an object, in $\leq 25$-years through the use of atmospheric drag, that would otherwise remain in orbit |

**Table 7      Needs matrix of a gossamer structure for end-of-life disposal**

## VI.  Conclusion

A gossamer structure is best suited to end-of-life disposal through atmospheric drag augmentation. This does not significantly alter the operating ceiling altitude of a spacecraft but does significantly alter the maximum allowable end-of-life mass. Peak mass benefit occurs in the altitude range 550 – 650 km and is largely independent of de-orbit time. The principal advantage of any atmospheric drag augmentation system for end-of-life disposal is the entirely passive operational mode, allowing the system to act as a 'fail-safe' that would activate if the spacecraft suffers a catastrophic failure.

## VII.  Acknowledgments